\documentclass[journal,10pt,twocolumn]{IEEEtran}
\usepackage{epsfig}
\usepackage{graphicx}
\usepackage{subfigure}

\usepackage{amssymb}
\usepackage{makeidx}
\usepackage{amsmath}
\usepackage{graphicx}
\usepackage{epsf}
\usepackage{epsfig}
\usepackage{psfig}
\usepackage{ccaption}
\usepackage{array}
\usepackage{tabularx}
\usepackage{multirow}
\usepackage{epsfig}
\usepackage{cite}
\usepackage{procedure,algorithm,algorithmic}
\begin{document}
%\ninept
%

\title{Need-based Communication for Smart Grid: When to Inquire Power Price?}
%
% Single address.
% ---------------
\author{Husheng Li and Robert C. Qiu
\thanks{H. Li is with the Department of Electrical Engineering
and Computer Science, the University of Tennessee, Knoxville, TN,
37996 (email: husheng@eecs.utk.edu). R. C. Qiu is with the Department
of Electrical and Computer Engineering, Tennessee Technological University, Cookeville, 38505 TN. This work was supported by the National Science
Foundation under grants CCF-0830451,  MRI-0821658 and
ECCS-0901425.}
}

\maketitle
\begin{abstract}
In smart grid, a home appliance can adjust its power consumption level according to the realtime power price obtained from communication channels. Most studies on smart grid do not consider the cost of communications which cannot be ignored in many situations. Therefore, the total cost in smart grid should be jointly optimized with the communication cost. In this paper, a probabilistic mechanism of locational margin price (LMP) is applied and a model for the stochastic evolution of the underlying load which determines the power price is proposed. Based on this framework of power price, the problem of determining when to inquire the power price is formulated as a Markov decision process and the corresponding elements, namely the action space, system state and reward function, are defined. Dynamic programming is then applied to obtain the optimal strategy. A simpler myopic approach is proposed by comparing the cost of communications and the penalty incurred by using the old value of power price. Numerical results show the significant performance gain of the optimal strategy of price inquiry, as well as the near-optimality of the myopic approach.
\end{abstract}
\section{Introduction}
In recent years, smart grid \cite{Smart2009}\cite{Moslehi2010}\cite{Wen2010} has attracted significant attention in the field of power systems, communications and networking. In a smart grid, power is delivered from power suppliers to home appliances with the aid of two-way communications. The price of power changes with time, subject to many random factors like congestion level and power generation. Home appliances can inquire the instantaneous price and decide the consumption level of power. For example, at midnight, the power load is usually low and thus the price is relatively low; therefore, air conditioner may set a higher temperature (suppose that it is in the winter) and enjoy the low power price.

Existing studies usually pay attention to only the cost of power consumption and ignore the cost of communications. However, communications for inquiring the price in smart grid could incur a nonnegligible cost. Therefore, it is necessary to consider the cost of communications and study the optimal policy of requesting communications for power price inquiry, such that the total cost of power consumption and communications is minimized under certain constraints. To the authors' best knowledge, no existing work has incorporated the cost of communication into the decision procedure in smart grid.

Due to the cost of communication, it may not be optimal to inquire the power price frequently. If the power price changes slowly, it may be better to use old power price to optimize the power consumption level. However, using old power price is also risky. For example, when the old power price is much higher than the current power price, the home appliance may use a lower power consumption, thus wasting the opportunity of low power price if it still uses the old price and does not inquire the new one. When the old power price is much lower than the current price, the policy of using old price will result in a high power consumption level, thus incurring significant penalty. Therefore, an optimal tradeoff should be found between the cost of communications and the penalty incurred by using the old value of power price.

In this paper, we study the decision problem of communication for inquiring the power price in order to minimize the total cost. A key issue is the prediction of power price based on the current obtained price. A model called {\em probabilistic locational marginal pricing (LMP) forecasting} \cite{Bo2009}\cite{Li2007} is applied to predict the distribution of the power price at a given future time. In this model, a curve is used to map from the true value of power load to the power price. A probabilistic model similar to Brownian motion is used to describe the distribution of load as a functional of the elapsed time since the latest inquiry of the power price.
\begin{figure}
  \centering
  \includegraphics[scale=0.55]{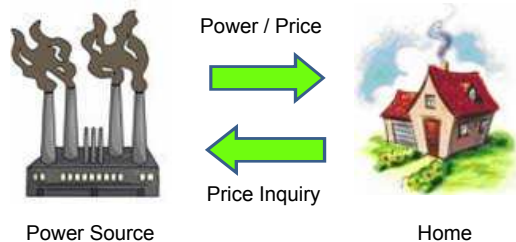}
  \caption{The simplified model of smart grid.}\label{fig:model}
\end{figure}
Once the evolution law of the power price (load) is known, we convert the problem into a Markov decision process (MDP) and obtain the corresponding state transition probabilities. Then, we apply dynamic programming to compute the optimal strategy. To simplify the strategy, we also propose a myopic strategy which compares the cost of communication with the penalty incurred by using the old power price. Note that we assume that the dynamics of the power price are perfectly known. For practical case, when there is no perfect model for the power price, we can apply the approach of reinforcement learning to learn the optimal strategy, which is beyond the scope of this paper.

The remainder of the paper is organized as follows. The system model and the price forecasting are introduced in Section \ref{sec:system}. The optimal communication policy is discussed in Section \ref{sec:inquiry}. Numerical results are provided in Section \ref{sec:numerical} while conclusions are drawn in Section \ref{sec:conclusion}.

\section{System Model and LMP Forecasting}\label{sec:system}
In this section, we first introduce the system model used in this paper. Then, we give a brief introduction to the pricing mechanism in the power grid, namely the LMP forecasting mechanism.

\subsection{System Model}
\begin{figure}
  \centering
  \includegraphics[scale=0.5]{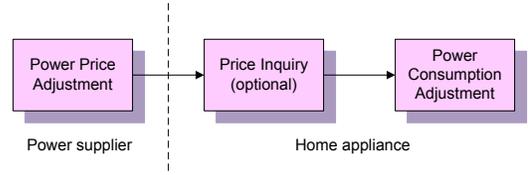}
  \caption{Events within a time slot.}\label{fig:timing}
\end{figure}

Practical power grid is very complicated. To simplify the analysis and facilitate the analytical discussion, we consider a simple supplier-home model, as illustrated in Fig. \ref{fig:model}. In this model, we consider only one power supplier and one home appliance. The power supplier provides power for the home appliance via power line, as well as the power price via a communication channel. The communication channel could be over Internet, wireless networks or power line communication systems. The home appliance inquires the power price via the communication channel. The following assumptions are used throughout the paper:
\begin{itemize}
\item We ignore possible errors over the communication channel and assume perfect communications. We do not consider the details of communications like modulation and coding.

\item Time is divided into time slots. At the beginning of each time slot, the power supplier adjusts its power price. Then, the home appliance can inquire the power price and receive the price information without or with delay. Once the price is obtained, the home appliance determines its power consumption level. These events are illustrated in Fig. \ref{fig:timing}.

\item For simplicity, we assume that the appliance can adjust its power consumption according to the power price immediately. In practice, there could be a delay for the power consumption adjustment, e.g. it needs some time to start the air conditioner. The corresponding analysis will be more complicated and is beyond the scope of this paper.
\end{itemize}

We denote by $p_t$ the power price and $x_t$ the power consumption at the $t$-th time slot. The utility function of the power consumption is denoted by $U(x_t)$, i.e. the home receives reward $U(x_t)$ when the power consumption is $x_t$. For simplicity, we assume that the utility function does not change with time. The cost for one communication effort is denoted by $c$, which is assumed to be a constant. We denote by $I_t$ the event that the home inquires the power price at the $t$-th time slot, i.e. $I_t=1$ if it inquires and $I_t=0$ otherwise. We denote by $\tau_t$ the latest time slot before time slot $t+1$ in which the home inquired the power price. The home appliance uses the same power price since the previous price inquiry, namely $p_{\tau_t}$ at time slot $t$. Therefore, the decision of power consumption is based on the power price of the previous price inquiry, i.e. $x_t$ is a function of $p_{\tau_t}$. For simplicity, we assume that the power consumption level maximizes the net reward and ignores the communication cost, i.e.
\begin{eqnarray}
x_t(p)=\arg\max_{x}\left(U(x)-px\right),
\end{eqnarray}
where $p$ is the price assumed by the home appliance (may be different from the true value if the home appliance does not inquire the power price). We assume that $U$ is an increasing and strictly concave function (thus the first order derivative is strictly decreasing). We also assume that $U$ is continuously differentiable and its first order derivative, denoted by $\dot{U}$ ranges from $\infty$ to $0$. Therefore, the optimal value of the power consumption level is given by
\begin{eqnarray}
x_t(p)=\dot{U}^{-1}(p),
\end{eqnarray}
which is derived from the first order condition, i.e.
\begin{eqnarray}\label{eq:price}
\dot{U}(x)-p=0.
\end{eqnarray}
Since $\dot{U}$ ranges from $\infty$ to 0 and is continuous and strictly decreasing, there exists a unique solution to (\ref{eq:price}). Hence, we have a one-to-one mapping between the price and the optimal power consumption level.

Although there are some simplifications in these quantities, we can obtain the insight from the simplified model and extend them to more generous case in the future.
Based on the above definitions, we assume that the total reward of the home appliance is the discounted sum of the rewards in different time slots, which is given by
\begin{eqnarray}
R=\sum_{t=0}^\infty \beta^t \left(U(x_t)-p_{\tau_t}x_t-cI_t\right),
\end{eqnarray}
where $\beta$ is the discount factor.

\subsection{LMP Forecasting}
\begin{figure}
  \centering
  \includegraphics[scale=0.5]{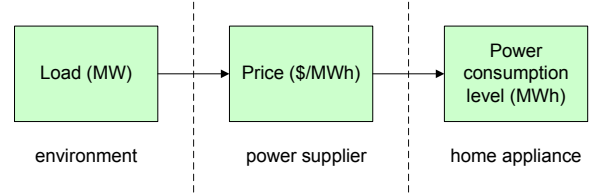}
  \caption{The dependency of factors in the supply chain.}\label{fig:factors}
\end{figure}
\begin{figure}
  \centering
  \includegraphics[scale=0.5]{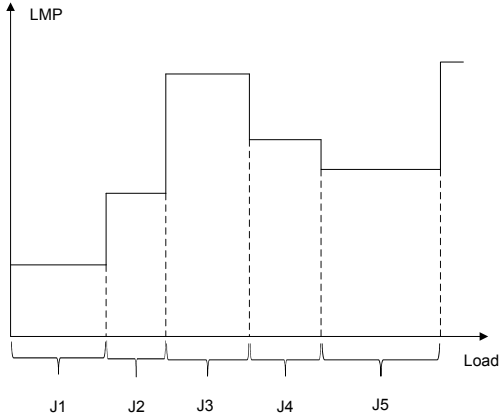}
  \caption{An illustration of the mapping between load and LMP.}\label{fig:mapping}
\end{figure}

Power price is usually determined by LMP methodology, which is actually driven by the time-varying load, as illustrated in Fig. \ref{fig:factors}. The mapping between load and LMP is typically obtained from a constrained optimization problem \cite{Stoft2002}. In practice, the mapping can be represented by a piecewise curve, as illustrated in Fig. \ref{fig:mapping}. Therefore, the uncertainty of the power price is from that of load. We notice that, in Fig. \ref{fig:mapping}, the number of possible prices is finite. Therefore, we denote by $K$ the total number of possible prices and by $q_1$, $q_2$, ..., $q_K$ the corresponding prices. Meanwhile, we denote by $J_1$, ..., $J_K$ the load intervals corresponding to the prices. Given a price $q_i$, we assume that the load is uniformly distributed within the interval $J_i$.

Actually the load is a random variable. Typically, it is modeled as a Gaussian random variable \cite{Bo2009}, i.e.
\begin{eqnarray}
D_t\sim \mathcal{N}(\mu_t,\sigma_t^2),
\end{eqnarray}
where $D_t$ is the load at time slot $t$, $\mu_t$ and $\sigma_t$ are the corresponding expectation and variance. Note that the assumption of Gaussian random variable is an approximation since a Gaussian random variable could be negative while a negative price is nonsense. To make it mathematically rigorous, we modify the probability density function (PDF) of $D_t$ as
\begin{eqnarray}\label{eq:evolution}
f(D_t)=\frac{\exp\left(-\frac{(D_t-\mu_t)^2}{2\sigma_t}\right)}{\int_0^{D_{\max}} \exp\left(-\frac{(y-\mu_t)^2}{2\sigma_t}\right)dy},
\end{eqnarray}
where $D_{\max}$ is the maximal possible load.

To study the optimal inquiry time for communications, we need to model the relationship between the time interval and the Gaussian distribution parameters. Suppose that, at time slot 0, the true value of the load, $D_0$, is known and then there is no observation on the true value of the load. Then, we use the following assumptions for the load distribution at time slot $t$:
\begin{itemize}
\item The expectation $\mu_t$ equals $D_0$, i.e. the expectation does not change with time, which represents an unbiased price prediction.
\item The variance $\sigma_t$ equals $\sigma_t=\theta t$, i.e. the variance increases linearly with the time gap, similarly to a Brownian motion. The parameter $\theta$ can be estimated from historical data. The rationale behind the Brownian motion like variance is that Brownian motion is widely used to drive the price fluctuation, e.g. stochastic differential equation, in the area of financial analysis \cite{Steele2000}. Therefore, we also use a linearly increasing variance to model the increasing uncertainty with time, which stimulates the price inquiry over the communication channel.
\end{itemize}

\section{Optimal Policy for Price Inquiry}\label{sec:inquiry}
In this section, we study the optimal policy for price inquiry over the communication channel. We can model the decisions of price inquiry as a Markov decision process (MDP) \cite{Bertsekas1987}. We will discuss the fundamental elements in the MDP. Then, we apply the approach of value iteration to obtain the optimal policy. Note that we assume that the statistical laws of the price, like the mapping between the load and the price and the uncertainty on the load, are all known to the home appliance.

\subsection{MDP Modeling}
There are three elements in a MDP problem, namely action space, system state and reward. We discuss them in the context of price inquiry separately.

\subsubsection{Action} Obviously, the action of the home appliance is the inquiry of the power price (denoted by 1) or not (denoted by 0). The decision of action is determined by the current system state and the policy of the home appliance. Actually, there is an implicit action for the home appliance, i.e. the power consumption level. In the most complicated case, the power consumption level should also be a function of the current system state and policy. However, to simplify the analysis, we assume that the home appliance assumes the power price of the latest inquiry, thus uniquely determines the power consumption level. Therefore, we do not consider the power consumption level as an action and do not incorporate it into the decision policy.

We put an upper bound, denoted by $T$, for the number of time slots between two price inquiries. Then, the home appliance must send out a price inquiry within $T$ time slots since the previous price inquiry. The upper bound can address the jeopardy brought by modeling imperfection, e.g. some imprecise parameters may significantly lengthen the interval between two inquiries to a harmful level, and thus improves the robustness.

\subsubsection{System State} 
\begin{figure}
  \centering
  \includegraphics[scale=0.55]{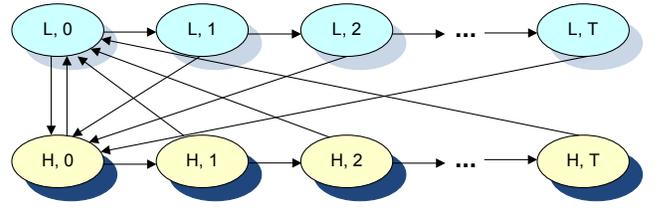}
  \caption{The diagram of state transition.}\label{fig:state}
\end{figure}

The system state contains two parts, namely the power price in the previous inquiry and the elapsed time since the previous inquiry. For time slot $t$, the system state is given by $(p_{\tau_t}, t-\tau_t)$. Obviously, if the home appliance inquires the power price at time slot $t$, the corresponding system state is $(p_t,0)$. An illustration for the state transition diagram is shown in Fig. \ref{fig:state} for the case of two possible power prices (high or low). We notice that, whenever the price substate is changed, the substate of elapsed time is reset to 0.
Another important issue in the system state is the state transition probability respect to the action. Obviously, if the action is no inquiry, i.e. 0, the state is changed to $(q_i,\Delta+1)$ if the previous state is $(q_i, \Delta)$. When the action is inquiry, i.e. 1, the state is changed to $(q_j,0)$ from the previous state $(q_i,\Delta)$, where $j$ is a random variable. We denote by $K_{ij}(\Delta)$ the probability of transiting from price $q_i$ to price $q_j$ after $\Delta$ time slots. The transition probability is then given by
\begin{eqnarray}\label{eq:transition}
K_{ij}(\Delta)&=&P(p_\Delta=q_j|p_0=q_i)\nonumber\\
&=&P(D_\Delta\in I_j|p_0=q_i)\nonumber\\
&=&\int_0^\infty P(D_\Delta\in I_j|D_0, p_0=q_i)dD_0\nonumber\\
&=&\frac{1}{|J_i|}\int_{J_i}P(D_\Delta\in I_j|D_0)dD_0\nonumber\\
&=&\frac{1}{|J_i|}\int_{J_i}\int_{J_j}f(D_\Delta|D_0)dD_\Delta dD_0,
\end{eqnarray}
where $|J_i|$ is the length of interval $J_j$ and $f(D_\Delta|D_0)$ is the conditional PDF of $D_\Delta$ given $D_0$. From (\ref{eq:evolution}) and the assumption on the expectation and variance, it is easy to verify that the conditional PDF is given by
\begin{eqnarray}
f(D_\Delta|D_0)=\frac{\exp\left(-\frac{(D_\Delta-D_0)^2}{2\Delta\theta}\right)}{\int_{0}^\infty\exp\left(-\frac{(y-D_0)^2}{2\Delta\theta}\right)dy}.
\end{eqnarray}
\subsubsection{Reward} Suppose that the system state of the previous time slot is $(q_i,\Delta)$. If the action of the current time slot is 0, i.e. no inquiry, the expected reward of the current time slot is given by (recall that the power consumption level $x$ is a function of the assumed price)
\begin{eqnarray}\label{eq:reward1}
r(0)=\sum_{j}K_{ij}(\Delta)(U(x(q_i))-q_jx(q_i)).
\end{eqnarray}
Otherwise, the reward is given by (recalled that $c$ is the cost of communication)
\begin{eqnarray}\label{eq:reward2}
r(1)=\sum_{j}K_{ij}(\Delta)(U(x(q_j))-q_jx(q_j))-c.
\end{eqnarray}

\subsection{Value Iteration}
Once defining the elements of MDP, we can apply Dynamic Programming (DP) \cite{Bertsekas1987} to obtain the optimal strategy. We denote by $R(s)$ the optimal expected total reward when the initial state is $s$. Then, $R(s)$ satisfies the following Bellman's equation \cite{Bellman1957}, which is given by
\begin{eqnarray}\label{eq:Bellman}
R(s)=\max_{a}\left(r(s,a)+\beta E_{s,a}\left[R(s')\right]\right),
\end{eqnarray}
where $r(s,a)$ is the expected reward due to action $a$ and system state $s$, $E_{s,a}$ is the expectation conditioned on $a$ and $s$, $s'$ is the system state in the next time slot. The expectation can be computed using the state transition probability in (\ref{eq:transition}). The instantaneous reward $r(s,a)$ can be computed using (\ref{eq:reward1}) and (\ref{eq:reward2}).

The Bellman's equation can be solved by using the following value iteration \cite{Bertsekas1987}, which is given by
\begin{eqnarray}
R^{(t)}(s)=\max_{a}\left(r(s,a)+\beta E_{s,a}\left[R^{(t-1)}(s')\right]\right),
\end{eqnarray}
where the superscript $t$ is the index of iteration. The iteration converges to the solution of the Bellman's equation as $t\rightarrow\infty$. The optimal action is obtained from
\begin{eqnarray}
a^*(s)=\arg\max_{a}\left(r(s,a)+\beta E_{s,a}\left[R(s')\right]\right).
\end{eqnarray}

\subsection{Myopic Strategy}
In the Bellman's equation (\ref{eq:Bellman}), the optimal action based on the current system state should take the future reward into account, thus making the solution complicated. A simpler but perhaps suboptimal scheme is the myopic strategy, i.e. optimizing the instantaneous reward $r$ without considering the future system state evolution. Then, the corresponding action is given by
\begin{eqnarray}
a=\left\{
\begin{array}{ll}
0, &\qquad \mbox{if }r(0)\geq r(1)\\
1, &\qquad \mbox{if }r(0)< r(1)
\end{array}
\right.,
\end{eqnarray}
i.e. do not inquire the power price if the reward of no inquiry is larger than that of inquiry, and vice versa. By applying the expressions of rewards in (\ref{eq:reward1}) and (\ref{eq:reward2}), we have
\begin{eqnarray}\label{eq:myopic}
a=\left\{
\begin{array}{ll}
0, &c\geq\sum_{j}K_{ij}(\Delta)\delta r_{ij} \\
1, &c<\sum_{j}K_{ij}(\Delta)\delta r_{ij}
\end{array}
\right.,
\end{eqnarray}
where $r_{ij}$ is defined as
\begin{eqnarray}
\delta r_{ij}=U(x(q_i))-U(x(q_j))-q_j(x(q_i)-x(q_j)),
\end{eqnarray}
which stands for the penalty of using the old value of the power price.
Then, the decision rule (\ref{eq:myopic}) means that, when the cost of communication is larger than the penalty of using the old power price instead of the new one, the action should be no price inquiry; and the home appliance should inquire the power price, otherwise.

\section{Numerical Results}\label{sec:numerical}
\begin{figure}
  \centering
  \includegraphics[scale=0.5]{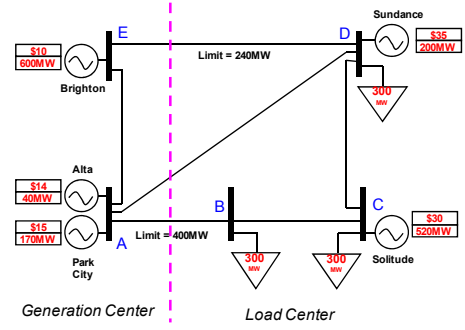}
  \caption{The base case modified from the PJM five-bus system \cite{PJM}.}\label{fig:PJM}
\end{figure}

\begin{figure}
  \centering
  \includegraphics[scale=0.5]{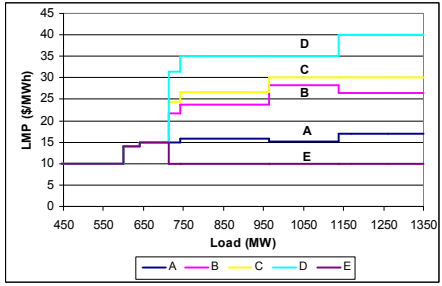}
  \caption{The LMP-load curve \cite{Li2007}.}\label{fig:curve}
\end{figure}

In this section, we use numerical simulations to explore the optimal power price inquiry based on the above framework. We use the PJM five-bus power system \cite{PJM} for simulations, whose configuration is illustrated in Fig. \ref{fig:PJM}. The corresponding curves of LMP versus load for the five buses are shown in Fig. \ref{fig:curve} and the lower boundaries for the load intervals are given in Table \ref{tab:curve} (note that $K=7$). Both the curves and the data are obtained from the continuous LMP model in \cite{Li2007}.

\begin{table}[ht]
\caption{LMP (\$/MWh) versus load (MW) \cite{Li2007}}
\label{tab:curve}\centering
\begin{tabular}{c c c c c c}
  \hline
  Load (MW) & LMP(A) & LMP(B) & LMP(C) & LMP(D) & LMP(E)  \\
  \hline
  0.00   & 10.00 & 10.00 & 10.00 & 10.00 & 10.00 \\
  \hline
  600.00 & 14.00 & 14.00 & 14.00 & 14.00 & 14.00 \\
  \hline
  640.00 & 15.00 & 15.00 & 15.00 & 15.00 & 15.00 \\
  \hline
  711.81 & 15.00 & 21.74 & 24.33 & 31.46 & 10.00 \\
  \hline
  742.80 & 15.83 & 23.68 & 26.70 & 35.00 & 10.00 \\
  \hline
  963.94 & 15.24 & 28.18 & 30.00 & 35.00 & 10.00 \\
  \hline
 1137.02 & 16.98 & 26.38 & 30.00 & 39.94 & 10.00 \\
  \hline
 1484.06 & 16.98 & 26.38 & 30.00 & 39.94 & 10.00 \\
  \hline
\end{tabular}
\end{table}

\subsection{Optimal Communication Strategy}
We assume that the upper bound of time slots between two inquiries is 10 time slots. We set $\beta=0.99$ for the discounted sum of cost as the performance metric. The utility function of the home appliance is assumed to be $U(x)=100\log x$. The default values of $\theta$ and $c$ are set to 200 and 10.

\begin{figure}
  \centering
  \includegraphics[scale=0.4]{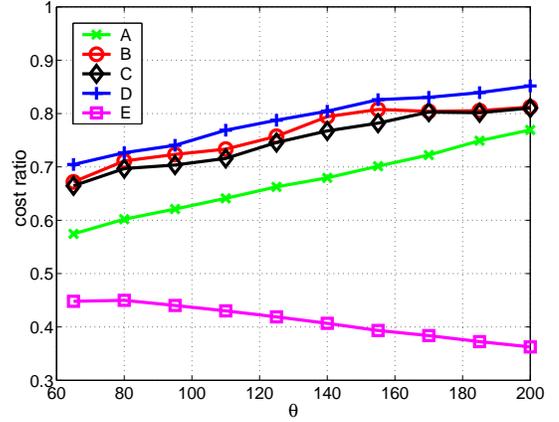}
  \caption{The ratio of cost between the optimal strategy and the always-inquiry strategy with different $\theta$.}\label{fig:cost_ratio}
\end{figure}

We first use the value iteration to obtain the optimal strategy and the corresponding minimum discounted sum of cost. Note that the cost here is defined as the gap of the corresponding reward to the ideal reward when there is no communication cost. We also tested the performance of the strategy of inquiring the power price in every time slot. The ratios of the discounted sum of cost obtained from the optimal strategy and that of the always-inquiry strategy are then computed for different values of $\theta$ and are shown in Fig. \ref{fig:cost_ratio}. There are five curves since there are five buses in the simulated power system. Obviously, the smaller the ratio is, the better the performance gain of the optimal strategy of price inquiry is. We observe that, for buses A, B and C, the ratio ranges between 0.6 and 0.8, i.e. the optimization can decrease the total cost by 20\% to 40\%. The performance gain of bus D is smaller. The reason is that the price changes the most radically for bus D, thus requiring more price inquiries. Since the power price of bus E has only marginal changes with respect to the change of load, it is much less necessary to inquire the price, thus making the performance gain of bus E much more significant than other buses.

\begin{figure}
  \centering
  \includegraphics[scale=0.4]{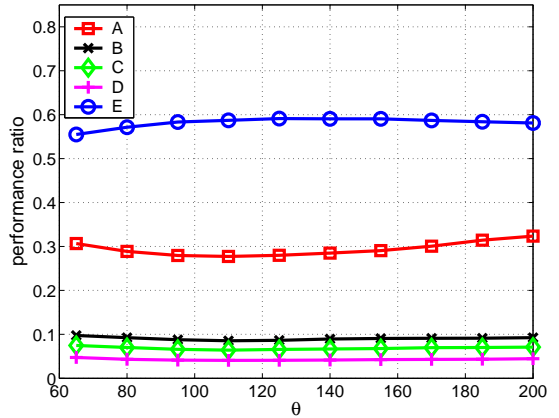}
  \caption{The ratio of cost between the optimal strategy and the no-inquiry strategy with different $\theta$.}\label{fig:cost_ratio2}
\end{figure}

Then, we compare the performance of the optimal strategy with that of the no-inquiry strategy. The curves of cost ratios are shown in Fig. \ref{fig:cost_ratio2}. In sharp contrast to Fig. \ref{fig:cost_ratio}, the order of the performance gain is reversed. The performance gain of bus E is the least since the necessity of price inquiry is the least for bus E. However, even for bus E, where the price changes only marginally, the optimal strategy can still achieve a very significant gain, thus demonstrating the necessity of price inquiry in smart grid.

\begin{figure}
  \centering
  \includegraphics[scale=0.4]{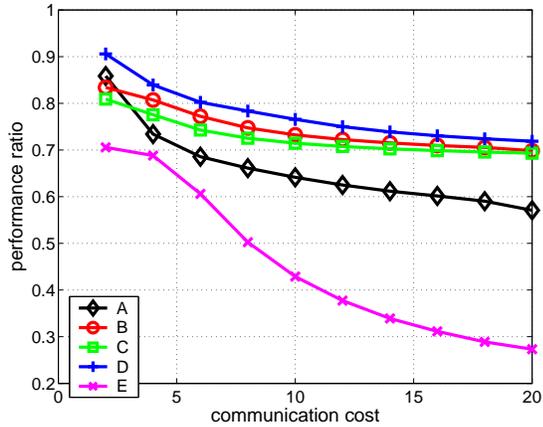}
  \caption{The ratio of cost between the optimal strategy and the always-inquiry strategy with different communication costs.}\label{fig:cost_ratio3}
\end{figure}

\begin{figure}
  \centering
  \includegraphics[scale=0.4]{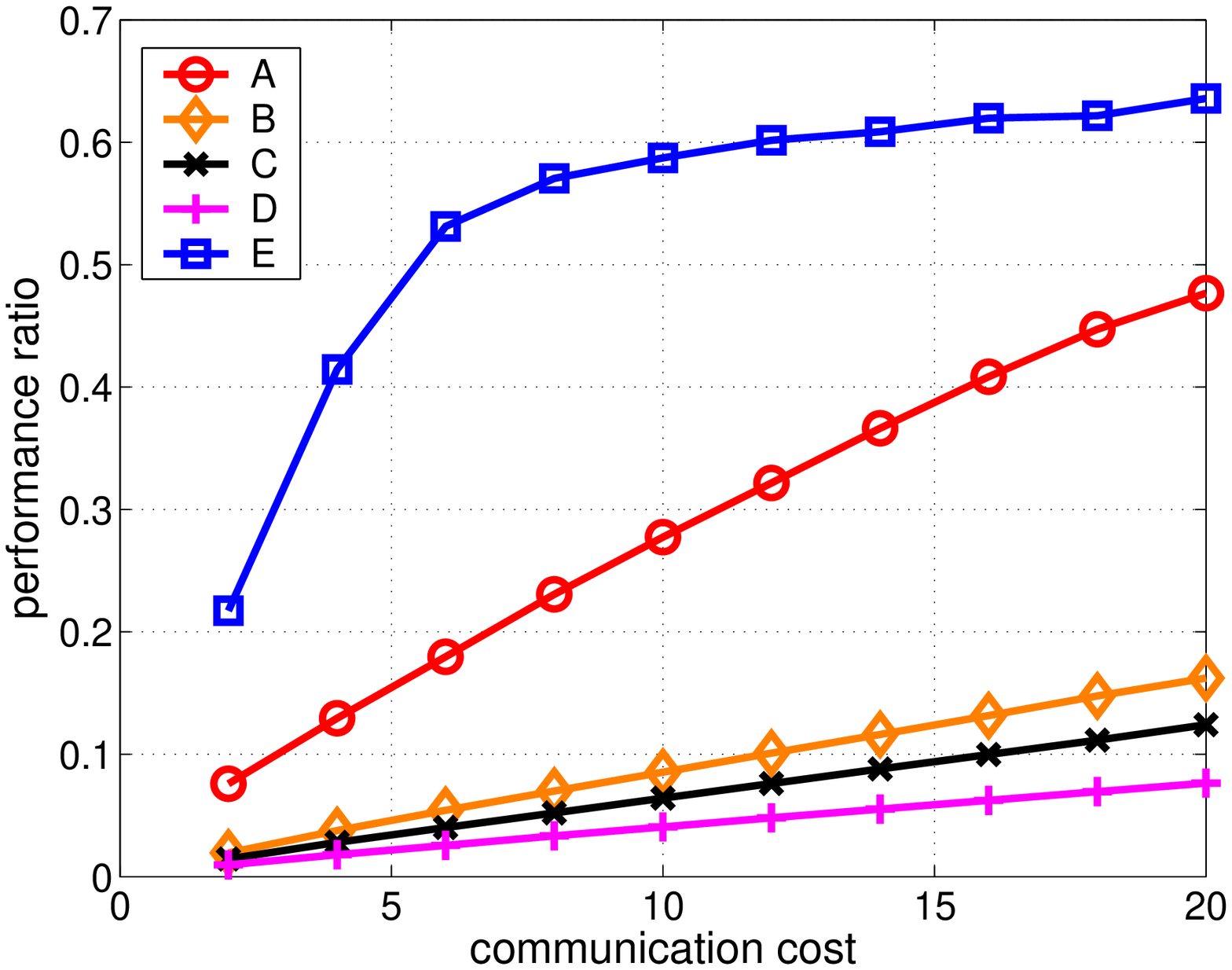}
  \caption{The ratio of cost between the optimal strategy and the no-inquiry strategy with different communication costs.}\label{fig:cost_ratio4}
\end{figure}

We repeated the simulations in Figures \ref{fig:cost_ratio} and \ref{fig:cost_ratio2} for different communication costs and fixed $\theta$. From both figures, the impacts of increasing communication cost on the performance gain are contrary. As the communication cost increases, the home appliance should be more inclined not to inquire the power price to reduce the cost of communication. Therefore, the performance gain in Fig. \ref{fig:cost_ratio3} increases while that in Fig. \ref{fig:cost_ratio4} is decreased.

\begin{figure}
  \centering
  \includegraphics[scale=0.4]{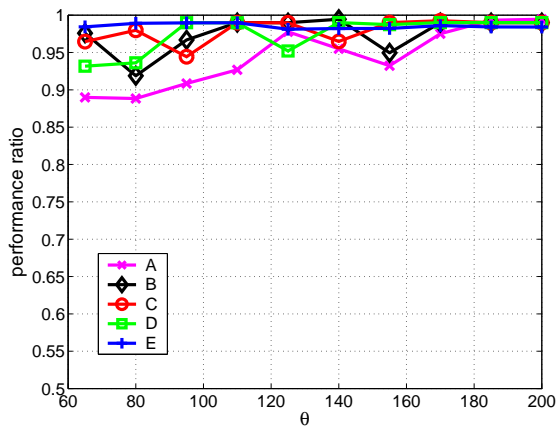}
  \caption{The ratio of cost between the optimal strategy and the myopic strategy with different $\theta$.}\label{fig:cost_ratio5}
\end{figure}

\begin{figure}
  \centering
  \includegraphics[scale=0.4]{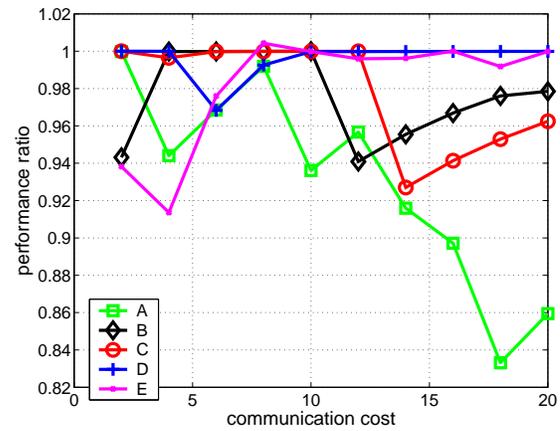}
  \caption{The ratio of cost between the optimal strategy and the myopic strategy with different communication costs.}\label{fig:cost_ratio6}
\end{figure}

\subsection{Myopic Strategy}
In Figures \ref{fig:cost_ratio5} and \ref{fig:cost_ratio6}, we compare the performance of the optimal strategy and the myopic strategy with different $\theta$ or different communication costs. In the ratio of cost, the numerator is the cost of the optimal strategy while the denominator is that of the myopic strategy. We observe that the myopic strategy is very close to optimal. Therefore, in practical systems, one may consider the myopic strategy due to its simplicity.

\section{Conclusions}\label{sec:conclusion}
We have considered the communication cost in smart grid, which is often omitted in existing studies. The dynamics of power price have been modeled as a Markov chain by modeling the random process of load as a Brownian motion like one and employing the LMP-load mapping curve. Then, the decision of power inquiry has been considered as a MDP problem and dynamic programming is employed to compute the optimal strategy. To avoid the high computational cost, we have studied a simple and suboptimal myopic strategy. A PJM five-bus system has been used for numerical simulation, which shows significant performance gain of the optimal strategy of price inquiry, as well as the near-optimality of the myopic approach.

% ------------------------------ BIBLIOGRAPHY ----------------------------------
\bibliographystyle{IEEE}

\end{document}